\documentclass[12pt]{article}

\usepackage[english]{babel}
\usepackage{amssymb,latexsym,amsmath,mathrsfs}

\begin{document}

\title
{Solitons and exotic instantons}
\author
{E.K. Loginov\footnote{{\it E-mail address:} ek.loginov@mail.ru}\\
\it Ivanovo State University, Ivanovo, 153025, Russia}
\date{}
\maketitle

\begin{abstract}
In this paper, we study the 't Hooft type instantons in eight dimensions, which satisfy the (anti)self-dual equations $F\wedge F=\pm\ast_8F\wedge F$. Using various designs of such instantons, we find new soliton solutions to the low-energy effective theory of the heterotic fivebrane. We investigate conditions under which these instanton configurations can be identified with the $D$-instantons embedded in the $D7$-brane world volume. Finally, we discuss the relationship between eight-dimensional periodic instantons and monopoles.
\end{abstract}

\section{Introduction}

The study of classical soliton solutions in string theory with higher brane structure is closely related to the construction of nonperturbative superstring theory. Such solitons are static multisoliton solutions that saturate the Bogomol'nyi bound. While classical solitons are three-level solutions in quantum string theory, they can nevertheless be used in nonperturbative calculations such as vacuum tunneling, since higher-order corrections often do not contribute to these effects.
\par
The first major breakthrough for $p$-brane solitons came with the paper~\cite{stro90}, where was showed that the ten-dimensional supergravity coupled to super Yang-Mills admits as a solution the heterotic fivebrane. The most important part of the design was the Yang-Mills instanton located in four directions transverse to the fivebrane. Later, the Yang-Mills instantons were used to construct classical $p$-brane solitonic solutions of the low energy effective theory of the heterotic string~(see~\cite{call91,harv91,khur93,guna95,logi05,logi08,ivan05,gemm13}).
\par
At the same time, there are solutions to the self-duality equations between the higher-order terms of the Yang-Mills field. This kind of Yang-Mills theory is not the standard gauge field theory but appears in some physically appropriate situations. For example, in~\cite{duff91,lu92}, the quartic Yang-Mills model has been used to find soliton solutions of the heterotic fivebrane. A crucial part of this construction was the one-instanton in eight dimensions satisfying the equations of self-duality $F\wedge F=\ast_8F\wedge F$. As shown in~\cite{duff97}, this instanton plays an essential role in smoothing out the singularity of heterotic string soliton solutions by incorporating one-loop corrections.
\par
An explicit example of one-instanton satisfying such relations was originally proposed in~\cite{gros84,gros89}, as a classical configuration that minimizes the quartic action $\int d^8x\,\text{tr}(F\wedge F)^2$. Later it was realized~\cite{mina01} that it also minimizes the quartic action which appears from the dimensional reduction of the $\alpha'$ corrected super Yang-Mills theory in ten-dimensions~\cite{berg01}. This made it possible to consider the one-instanton as the $D$-instanton, i.e as an instanton embedded in the $D7$-brane. It was shown in~\cite{bill09a} that the one-instanton solution can be interpreted as the $D$-instantons if their size becomes zero. A similar statement for the multi-instantons in the well-separated limit was obtained in~\cite{naka16}. The situation where the instanton solutions can be interpreted as the $D$-instantons has been analyzed in~\cite{bill09,bill21} where the instanton partition function for the $D7/D(-1)$ system was studied.
\par
In this article, we investigate various constructions of (anti)self-dual instantons in eight dimensions and seek related to them soliton solutions to the low-energy effective theory of the heterotic fivebrane. In the next section, we will study the self-dual two-instanton configurations of the 't Hooft type. We use such instantons to construct soliton solutions that generalize the fivebrane solution that was obtained in the case of one-instanton. In Sec.~III, we will find a solution to the eight-dimensional antiself-duality equations, and then show that this solution can be extended to larger dimensions.
In Sec.~IV, we will investigate the instanton effects of the obtained solutions. We will show that self-dual two-instanton configurations satisfy the $D$-instanton conditions in the zero-size and can be identified with the $D$-instantons embedded in the $D7$-brane world-volume. In Sec.~V, we will discuss the eight-dimensional periodic instantons (calorons) and their monopole limits. In Sec.~VI,  we will briefly list the results obtained.

\section{Multifivebrane configurations}

As mentioned in the Introduction, the ten-dimensional supergravity coupled to super Yang-Mills admits the fivebrane soliton as a solution, with the self-dual one-instanton being the key part of the design. In~\cite{duff91} (see also~\cite{duff95}) it was suggested that the heterotic string and the heterotic fivebrane may be dual to each other in the sense that they are equivalent descriptions of the same underlying physical theory. Under this assumption, the heterotic fivebrane action takes the form
\begin{equation}\label{12}
S=\frac{1}{2\kappa^2}\int d^{10}x\,\sqrt{-g}\left(R-\frac{1}{2}(\partial{\phi})^2-\frac{1}{2\cdot7!}e^{\phi}K^2
-\frac{\beta'}{2^44!}e^{\phi/2}\text{tr}(t_8F^4)\right).
\end{equation}
Here $K$ is the seven-form field strength, which is dual to the three-form field strength of the heterotic string, the fivebrane tension $T_6$ is given by $1/\beta'=(2\pi)^3T_6$, and $t_8$ is the ten-dimensional extension of the eight-dimensional light-cone gauge ``zero-mode'' tensor, i.e.
\begin{align}
t_8F^4&=F^{MN}F_{PN}F_{MS}F^{PS}+\frac{1}{2}F^{MN}F_{PN}F^{PS}F_{MS}\notag\\
&-\frac{1}{4}F^{MN}F_{MN}F^{PS}F_{PS}-\frac{1}{8}F^{MN}F^{PS}F_{MN}F_{PS}.
\end{align}
In this section, we use the self-dual two-instantons to construct fivebrane soliton solutions of the equations of motion that follow from the action (\ref{12}).
\par
Following~\cite{duff91,dabh90}, we split the indices $x^M=(x^{m},y^{\mu})$, where $m,n=1,2,\dots,8$ and $\mu,\nu=10,11$, and we choose a supergravity ansatz for the ten-dimensional metric $g_{MN}$ and seven-form $K_{MNPSIJKL}$ in the form
\begin{align}
&ds^2=e^{3\phi/2}\eta_{\mu\nu}dy^{\mu}dy^{\nu}+e^{-\phi/2}\delta_{mn}dx^mdx^n,\label{09}\\
&K_{mnpsijk}=2\sqrt{g_8}e^{-\phi/2}\varepsilon_{mnpsijkl}\partial^l\phi,\label{10}
\end{align}
where $g_8=\det g_{mn}=e^{-4\phi}$ and the dilaton field $\phi$ depends only on $x^m$. All other components of $K$ are set zero. With this choice of $g_{MN}$ and $K$, the supersymmetry variations $\delta\psi_M$ and $\delta\lambda$ of the gravitino and gaugino vanish.
\par
In order to make the variation $\delta\chi$ of the gaugino zero, we must make an ansatz for $A_M$ which preserves both the bosonic and fermionic symmetries. This configuration is provided by the (anti)self-dual $SO(8)$-instanton that is a solution of the equations
\begin{equation}\label{01}
F_{[mn}F_{ps]}=\pm\frac{1}{24}\varepsilon_{mnpsijkl}F_{[ij}F_{kl]}
\end{equation}
in the eight-dimensional Euclidean space with the flat metric. Here, the gauge field strength
\begin{equation}
F_{mn}=\partial_mA_n-\partial_nA_m+[A_m,A_n],
\end{equation}
where the potential $A_m$ taking value in the Lie algebra $so(8)$. Given the known explicit formula for the potential
\begin{equation}\label{19}
\tilde{A}_m=-\frac{R_{mn}x_n}{\lambda^2+x^2},
\end{equation}
where $R_{mn}=\frac{1}{2}(1+\Gamma_9)\Gamma_{mn}$ are generators of $SO(8)$ corresponding to the irreducible 8-dimensional representation of $Spin(8)$, and the gauge field strength
\begin{equation}\label{23}
\tilde{F}_{mn}=\frac{2\lambda^2R_{mn}}{(\lambda^2+x^2)^2}
\end{equation}
corresponding to a single instanton of the size $\lambda$, the self-duality equations (\ref{01}) can be solved (see~\cite{gros84}).
\par
We will look at a more general multi-instanton configuration with the potential
\begin{equation}\label{13}
A_{m}=-\frac{1}{2}R_{mn}\partial_{n}\ln\left(1+\sum^{N}_{i=1}\frac{\lambda_i^2}{(x-b_i)^2}\right),
\end{equation}
where $\lambda_i$ gives the scale size of the instanton located at $b_i$. As was shown in~\cite{logi20}, this potential defines the field strength that is a solution to the self-dual equation (\ref{01}). This multi-instanton can be viewed as a superposition of $N$ instantons that depend on $8N+1$ free parameters and its topological charge is defined by the fourth Chern number. To determine the dilaton field, we substitute $F_{mn}$ into the field equations of the fivebrane obtained by varying the action (\ref{12}). They are all satisfied identically, except for the dilaton field equation which reduces to
\begin{equation}\label{04}
\partial^m\partial_me^{-2\phi}=-\frac{2\cdot8!\beta'}{4!\Phi^4}\text{tr}f^4,
\end{equation}
where the function $\phi=\phi(x^{m})$ is identified with the dilaton field, and $f=f(x_i)$ are given by
\begin{equation}\label{03}
f=\left(\frac{\lambda_1x_1^t}{x_1^2}\dots\frac{\lambda_kx_k^t}{x_k^2}\right)
\begin{pmatrix}\lambda_1^2+x_1^2&\hdots&\lambda_1\lambda_k\\
\vdots&\ddots&\vdots\\
\lambda_1\lambda_k&\hdots&\lambda_k^2+x_k^2
\end{pmatrix}^{-1}
\begin{pmatrix}\frac{\lambda_1x_1}{x_1^2}\\\vdots\\\frac{\lambda_kx_k}{x_k^2}
\end{pmatrix}.
\end{equation}
Here $x_i=(x^m-b_i^m)R_m$, where $R_8\equiv I_8$ is the unit $8\times8$ matrix and $R_m$ is an image of $\Gamma_m$ as $m\ne8$ with respect to the homomorphism $Spin(8)\to SO(8)$, $x_i^t$ is the matrix transposed to $x_i$, and $\Phi=1+\sum^{N}_{i=1}\lambda_i^2/x_i^2$, where $x_i^2=(x^m-b_i^m)(x_m-b_{im})$.
\par
In the particular case, when $N =1$, a solution of (\ref{04}) can be obtained if looking for it in the form $\phi=\phi(x^2)$. In this case, one is transformed into an ordinary differential equation and we get the following dilaton solution
\begin{equation}\label{07}
e^{-2\phi}=1+\frac{c}{x^6}-16\beta'\frac{\lambda^8(15x^4+6\lambda^2x^2+\lambda^4)}{x^6(\lambda^2+x^2)^6},
\end{equation}
where $c$ is an integration constant and for convenience we have set $\phi_0=0$. If we make the identification $c=16\beta'$, then this solution will exactly coincide with the solution that was obtained in~\cite{duff91} for $SO(8)\subset SO(32)$. If we put $\beta'=0$ and $c>0$, then we obtain the solution of~\cite{lu92} for $SO(8)\subset E_8$.
\par
Now we suppose $N=2$. It follows from (\ref{03}) that
\begin{equation}\label{17}
\partial^m\partial_me^{-2\phi}
=-\beta'\frac{8!}{4!}\left[2\frac{\lambda_1^2x_2^4+\lambda_2^2x_1^4+\lambda_1^2\lambda_2^2(x_1^2+x_2^2-2x_1^{m}x_2^{m})}
{(\lambda_1^2x_2^2+\lambda_2^2x_1^2+x_1^2x_2^2)^2}\right]^4,
\end{equation}
where the parameters $\lambda_i$ determine the sizes of the instantons, and $x_i=x+b_i$ are their positions. It is convenient to pass to a new variable $\tilde{x}=x+\frac{1}{2}(b_2+b_1)$ and put $b=\frac{1}{2}(b_2-b_1)$. (For $b_2=b_1$, we obtain the dilaton solution (\ref{07}) with $\lambda^2=\lambda_1^2+\lambda_2^2$.) In this case, $x_1=\tilde{x}-b$ and $x_2=\tilde{x}+b$. Consequently, the origin moves to the middle of the segment connecting the instantons. In addition, using an appropriate orthogonal transformation, all but one component of $b$ can be nullified. Therefore, we can assume that $b=b^0I_8$.
\par
We will find solutions to Eq.~(\ref{17}) in the two limiting cases, when $|\tilde{x}|\gg |b|$ and $|\tilde{x}|\ll |b|$. In the first case, we again get the dilaton solution (\ref{07}). Therefore the large-distance behavior of the two-fivebrane is the same as that of the one-fivebrane The alternative condition $|\tilde{x}|\ll |b|$ leads to the following dilaton solutions
\begin{align}
e^{-2\phi}&=1-105\beta'\frac{2^7(\lambda^4\tilde{x}^0)^2}{b^8(\lambda^2+b^2)^4},\label{11}\\
e^{-2\phi}&=1-\beta'\frac{2^9(56z^0+5)}{(\lambda_2^2-\lambda_1^2)^2(2z^0+1)^6}\label{08}
\end{align}
as $|\lambda_2|=|\lambda_1|$ and $|\lambda_2|\ne|\lambda_1|$ respectively. Here $z^0=(\lambda_2^2-\lambda_1^2)\tilde{x}^0/|b|^3$ and $\lambda^2=\lambda_1^2+\lambda_2^2$. Note that the solution (\ref{08}) was found in the well-separated limit (see~\cite{chri78,naka16}). In this case the scale parameters $\lambda_1$ and $\lambda_2$ are of small order compared to the ``instanton separations'' $|x_1-x_2|$ and therefore the off-diagonal elements of the matrix (\ref{03}) can be neglected. Generally speaking, this condition is optional. However, if we do not ignore it, then we will get a very cumbersome expression. Thus, the ansatz (\ref{09}), (\ref{10}), (\ref{13}) with the dilaton fields (\ref{11}) or (\ref{08}) gives a multisoliton solutions of the fivebrane field equations.

\section{Higher-dimensional instantons}

Let us move on to finding the instantons of the antiself-dual configuration. First of all, note that any totally antisymmetric eight-dimensional tensor of the fourth rank can be written as the sum of the self-dual and the antiself-dual parts $F_{mnps}=F^{+}_{mnps}+F^{-}_{mnps}$, where
\begin{equation}\label{14}
F_{mnps}^{\pm}=\left(\delta^i_{[m}\delta^j_n\delta^k_p\delta^l_{s]}\pm\frac{1}{24}\varepsilon_{mnpsijkl}\right)F_{ijkl}.
\end{equation}
On the other hand, the identity $\Gamma_m\Gamma_n=\Gamma_{mn}-\delta_{mn}$ holds for the gamma matrices. (Here and below, we ignore the identity matrix, writing $\delta_{mn}$ instead of $\delta_{mn}I_8$.) Using it as a base of induction, it is easy to prove that
\begin{equation}\label{02}
\Gamma_{p}\Gamma_{s_1\dots s_k}=\Gamma_{ps_1\dots s_k}+\sum^k_{i=1}(-1)^i\delta_{ps_i}\Gamma_{ps_1\dots\hat{s}_{i}\dots s_k}.
\end{equation}
Let us denote again the generators of $SO(8)$ by $R_{mn}=\frac{1}{2}(1+\Gamma_9)\Gamma_{mn}$ and, using (\ref{02}), we rewrite the expression (\ref{14}) for the self-dual tensor in the following form
\begin{equation}
F_{mnps}^{+}=\frac{1}{24}\text{tr}\,(R_{mn}R_{ps}R_{ij}R_{kl})F_{ijkl}.
\end{equation}
Note that the last equality can be obtained in an alternative way if we use the identity
\begin{equation}
\Gamma_{i_1\dots i_k}=\frac{1}{(8-k)!}\varepsilon_{i_1\dots i_8}\Gamma_9\Gamma^{i_8\dots i_{k+1}}.
\end{equation}
Thus, the tensor $F_{mnps}$ is antiself-dual if $R_{mn}R_{ps}F_{mnps}=0$. In particular, the tensor $F_{mnps}=F_{[mn}F_{ps]}$ is antiself-dual if $R_{mn}F_{mn}=0$.
\par
Following~\cite{logi21}, we choose the ansatz
\begin{equation}\label{05}
A_m=-\frac{1}{6}R_{mp}\partial_{p}\ln\left(1+\frac{\lambda^2}{x^6}\right),
\end{equation}
where $\lambda$ gives the scale size of the instanton located at the origin. In order to find the gauge field strength $F_{mn}$, we use the identity
\begin{equation}\label{06}
R_{mp}R_{ns}=R_{mpns}+\delta_{mn}R_{ps}-\delta_{ms}R_{pn}-\delta_{pn}R_{ms}+\delta_{ps}R_{mn}
+\delta_{ms}\delta_{pn}-\delta_{mn}\delta_{ps},
\end{equation}
which is a consequence of (\ref{02}). As a result, we get
\begin{equation}\label{15}
F_{mn}=\frac{2\lambda^2x^2}{(\lambda^2+x^6)^2}(4R_{mp}x_nx_p-4R_{np}x_mx_p-R_{mn}x^2).
\end{equation}
Finally, using the identities
\begin{equation}\label{18}
R_{mn}R_{sn}=\delta_{nn}(R_{ms}-\delta_{ms}),\quad R_{mn}R_{mn}=-\delta_{mm}\delta_{nn},
\end{equation}
we prove the antiself-duality condition $R_{mn}F_{mn}=0$. Thus the potential (\ref{05}) is a solution of the antiself-duality equations. Note that the potentials (\ref{19}) and (\ref{05}) are gauge nonequivalent. To show this, it suffices to note that $\text{tr}F^2_{mn}\ne\text{tr}\tilde{F}^2_{mn}$.
\par
We now show that this eight-dimensional construction of the antiself-dual instanton can be generalized to large dimensions. In $d$-dimensional Euclidean space with $d>2$, we consider the gauge field potential
\begin{equation}\label{25}
A_m=-\frac{1}{d-2}R_{mp}\partial_{p}\ln\left(1+\frac{\lambda^2}{x^{d-2}}\right),
\end{equation}
where $R_{mp}$ are generators of $SO(d)$. To find the gauge field strength $F_{mn}$, we use the identity (\ref{06}). As a result, we get
\begin{equation}\label{26}
F_{mn}=\frac{\lambda^2x^{d-6}}{(\lambda^2+x^{d-2})^2}[d(R_{mp}x_n-R_{np}x_m)x_p-2R_{mn}x^2].
\end{equation}
Obviously, the fields (\ref{25}) and (\ref{26}) coincide identically with (\ref{05}) and (\ref{15}) as $d=8$. Using (\ref{18}), we obtain the identities
\begin{equation}\label{27}
R_{mn}F_{mn}=0,\quad F_{mnps}^2=0.
\end{equation}
Finally, we can find conditions under which this ansatz satisfies the Yang-Mills equations. In the considered dimension, we have
\begin{equation}
\partial_mF_{mn}+[A_m,F_{mn}]=(d-4)(1-d)\frac{4\lambda^4x^{d-6}}{(\lambda^2+x^{d-2})^3}R_{mp}x_p.
\end{equation}
Obviously, the required condition is satisfied only for $d=4$. In what follows, we will assume that the dimension $d=4n$, where $n\in\mathbb{N}$.
\par
In $4n$ dimensions, we can consider the (anti)self-dual equation
\begin{equation}
\ast F\wedge F\wedge\dots\wedge F=\pm F\wedge F\wedge\dots\wedge F,
\end{equation}
where the $2n$-forms are on both sides of the equality. It follows from the first equality in (\ref{27}) that the field strength $F_{mn}$, defined as (\ref{26}), is a solution of the antiself-dual equation. It follows from the second equality in (\ref{27}) that the $2n$-th Chern number
\begin{equation}
\frac{1}{(2n)!(2\pi)^{2n}}\int\text{tr}(F\wedge F\wedge\dots\wedge F)=0.
\end{equation}
If $n=1$, then the gauge fields take values in the Lie algebra $so(4)\simeq so(3)^{+}\oplus so(3)^{-}$. Suppose $A_m=A_m^{+}+A_m^{-}$, where $A_m^{\pm}\in so(3)^{\pm}$. Then the corresponding field strengths $F_{mn}^{\pm}$ become the (anti) self-dual Yang-Mills instantons, and their sum is $F_{mn}$. Therefore the second Chern number $k=k^{+}+k^{-}$ with $k^{\pm}=\pm1$. If $n>1$, then the gauge fields take values in the Lie algebra $so(4n)$. Since this algebra is simple, it is impossible to represent the $2n$-th Chern number as the sum $k=k^{+}+k^{-}$ with $k^{\pm}\ne0$.

\section{D-instantons}

We now consider the Euclidean $D7$-brane in Type IIB string theory. On the world-volume of this $D7$-brane, there is an eight-dimensional Yang-Mills theory which is naturally realized as low-energy effective field theory. In order to see the (anti)self-dual instanton effects of the obtained solutions, we consider the $\alpha'$ corrections to the gauge theory. The  gauge part of the effective action can be written as
\begin{equation}\label{22}
S_{D}=S_2+S_4+\dots.
\end{equation}
Here the first term is the quadratic Yang-Mills action in eight dimensions
\begin{equation}
S_2=\frac{1}{2g_{YM}^2}\int d^8x\,\text{tr}(F^2)
\end{equation}
with a dimensionful gauge coupling constant
\begin{equation}
g_{YM}^2=4\pi g_s(2\pi\sqrt{\alpha'})^4,
\end{equation}
where $g_s$ is the string coupling and $\sqrt{\alpha'}$ is the string length. The second term is a quartic action of the form
\begin{equation}
S_4=\frac{(4\pi\alpha')^2}{4!g_{YM}^2}\int d^8x\,\text{tr}(t_8F^4)-2\pi iC_0k,
\end{equation}
where $k$ is the fourth Chern number
\begin{equation}\label{24}
k=\frac{1}{4!(2\pi)^4}\int\text{tr}(F\wedge F\wedge F\wedge F)
\end{equation}
and $C_0$ is a scalar field of the closed string RR sector.
\par
Following~\cite{bill09a}, we will interpret the eight-dimensional instantons as the $D$-instantons, i.e. as instantons embedded in $D7$-branes. Such instantons are sources for RR 0-form $C_0$. In the case, the (anti)self-duality condition (\ref{01}) holds, the trace
\begin{equation}
\text{tr}(t_8F^4)=\pm\frac{1}{2}\text{tr}(F\wedge F\wedge F\wedge F),
\end{equation}
and hence the quartic action $S_4$ becomes
\begin{equation}
S_4=-2\pi i(C_0\pm\frac{i}{g_s})k.
\end{equation}
This precisely matches the action of the action of $k$ $D$-instantons. Thus, the eight-dimensional (anti)self-dual instantons become the $D$-instantons when $S_2=0$ and all the $O(\alpha'^4/g_{YM}^2)$ terms vanish. The second condition is fulfilled in the zero-slope limit $\alpha'\to0$ with fixed $\alpha'^2/g_{YM}^2$. In this limit, the quartic term remains finite while the Yang-Mills part diverges in general. In some cases, the condition $S_2=0$ is fulfilled in the zero-size limit $\lambda_i\to0$, i.e. when the sizes of instantons tend to zero. This has been shown for the self-dual one-instanton in~\cite{bill09a} and for the 't Hooft type multi-instantons in the well-separated limit in~\cite{naka16}. We will estimate the Yang-Mills quadratic term for the instantons that were above used to construct the fivebrane solitons.
\par
In the case of the two-instanton configuration with the potential (\ref{13}), the Yang-Mills quadratic term takes the form
\begin{equation}\label{20}
S_2=\frac{1}{g_{YM}^2}\int d^8x\left[4\frac{\lambda_1^2x_2^4+\lambda_2^2x_1^4+\lambda_1^2\lambda_2^2(x_1^2+x_2^2-2x_1^{m}x_2^{m})}
{(\lambda_1^2x_2^2+\lambda_2^2x_1^2+x_1^2x_2^2)^2}\right]^2,
\end{equation}
where $\lambda_i$ are the sizes of the instantons and $x_i=x+b_i$. We will calculate this integral in the two limiting cases, when $|\tilde{x}|\gg |b|$ and $|\tilde{x}|\ll |b|$, where as above $\tilde{x}=x+\frac{1}{2}(b_2+b_1)$ and $b=\frac{1}{2}(b_2-b_1)=b^0I_8$. If $|\tilde{x}|\gg |b|$, then we get the self-dual one-instanton of the size $\lambda=\sqrt{\lambda_1^2+\lambda_2^2}$. It then follows from~\cite{bill09a} that $S_2=0$ when the instantons shrink to zero-size $\lambda\to0$. Suppose $|\tilde{x}|\ll |b|$. Then the radial part of the space-time integral in (\ref{20}) can been rewritten in the form
\begin{equation}\label{21}
\int_0^R d^6r\left[\frac{4b^2r^3(\lambda^2(b^2+r)+4\lambda^2_{-}bz+4\lambda_1^2\lambda_2^2)}
{(\lambda^2(b^2+r)+2\lambda^2_{-}b\tilde{x}_0+b^2(b^2+2r-4\tilde{x}_0^2))^2}\right]^2
=\frac{8R^2b^4\lambda^4}{(2b^2+\lambda^2)^4},
\end{equation}
where $\lambda^2_{-}=\lambda_2^2-\lambda_{1}^2$. Here we introduce the cutoff $R$ in the space-time integral (\ref{20}) and neglect subleading terms of $1/R$. Obviously, when the instantons shrink to the zero-size limit, the Yang-Mills quadratic term vanishes. Thus, all the self-dual two-instantons, that we were used to constructing the fivebrane solitons, in the small instanton limit correspond to the $D$-instantons embedded in the $D7$-branes.
\par
Now we consider the anti-instanton configuration with the potential (\ref{05}). In this case, the Yang-Mills quadratic term has the form
\begin{equation}
S_2=-\frac{112^2}{g_{YM}^2}\int d^8x\frac{\lambda^4x^8}{(\lambda^2+x^6)^4}.
\end{equation}
The radial part of the space-time integral in (\ref{20}) has the form
\begin{equation}
\int_0^R d^7r\frac{\lambda^4r^{15}}{(\lambda^2+r^6)^4}
=\frac{\sqrt{3}\lambda^{4/3}}{3^7}\left(5\pi-30\arctan\frac{\lambda^{2/3}-2R^2}{3\lambda^{2/3}}\right),
\end{equation}
where we reintroduced the cutoff $R$ in the space-time integral and neglected the subleading terms $1/R$. Obvious that the Yang-Mills quadratic term $S_2$ vanishes when the instanton shrinks to the zero-size limit. However, this instanton has nothing to do with $D$-instantons embedded in the $D7$-branes. The point is that it satisfies the equality $\text{tr}F_{mnps}^2=0$ and therefore
\begin{equation}\label{16}
\frac{1}{12}\text{tr}(t_8F^4)=\frac{1}{4!\cdot 2^4}\text{tr}(\epsilon^{mnpsijkl}F_{mn}F_{ps}F_{ij}F_{kl})=0.
\end{equation}
Hence the quartic action in (\ref{22}) equal to zero. It follows from (\ref{16}) that the dilaton field corresponding to the antiself-dual instanton (\ref{15}) is given by the formula
\begin{equation}
e^{-2\phi}=1+\frac{c}{x^6},
\end{equation}
where $\phi_0=0$ and $c\geq0$. This formula coincides with (\ref{07}) as $\beta'=0$,  as well as with the singular elementary string solution obtained in~\cite{dabh90} (see also~\cite{duff95}).

\section{Calorons and monopoles}

In this section, we study the eight-dimensional periodic instantons (calorons) and their relationship with monopoles. In four dimensions, such instanton was first constructed in~\cite{harr79} as a finite temperature generalization of the BPST instanton. In~\cite{ross79,chak87}, a connection was found between calorons and magnetic monopoles, and in~\cite{gaun93}, discussed the relation between magnetic monopoles and fivebrane solutions. The higher-dimensional calorons were studied in~\cite{take17} but only in the well-separated limit.
\par
We will be interested in instanton solutions that are periodic in $t=x^0$. Such instanton can be constructed if we take an infinite string of identical instantons and will look for solutions to the self-dual equations (\ref{01}) on the space $R^7\times S^1$. Following~\cite{harr79}, we choose the 't Hooft-like ansatz (\ref{13}) and denote
\begin{equation}
\rho(x)=1+\sum^{N}_{i=1}\frac{\lambda_i^2}{(x-b_i)^2}.
\end{equation}
We suppose that all instantons have same identical gauge orientation (i.e. $\lambda_i=\lambda$ and $b_i=b$) and perform the summation to ensure the periodicity $t=t+2\pi R$. As a result, we get the eight-dimensional periodic instanton:
\begin{align}\label{29}
\rho(x)&=1+\sum^{\infty}_{k=-\infty}\frac{\lambda^2}{(t+2\pi kR)^2+r^2}\notag\\
&=1+\frac{\lambda^2}{2rR}\left(\frac{\sinh\frac{r}{R}}{\cosh\frac{r}{R}-\cos\frac{t}{R}}\right),
\end{align}
where $r=|\vec{x}|$ and the instanton is at the origin (we choose $b=0$). Under $\lambda\ll R$, this eight-dimensional instanton coincides with the one constructed in~\cite{take17}. However, the results obtained in~\cite{logi20} allow us to ignore this condition.
\par
The built instanton has two scale parameters, the instanton size $\lambda$ and the compactification radius $R$. The gauge fields of such instanton depend on the ratio of these scales. We will find the values of these fields in two limiting cases: at $|x|\ll R$ and at $|x|\gg R$. For distances $|x|\ll R$, the formula (\ref{29}) takes the form
\begin{equation}
\rho(x)=1+\frac{\lambda^2}{x^2}+\frac{\lambda^2}{R^2}O\left(\frac{x^2}{R^2}\right).
\end{equation}
In this case, the potential (\ref{13}) reduces to
\begin{equation}
A_m\sim\frac{R_{mn}x^n}{x^2(\lambda^2+x^2)}.
\end{equation}
This potential is gauge equivalent to (\ref{19}). Therefore, if we consider the scales much less than $R$, the eight-dimensional finite-temperature instanton is identical to a zero-temperature instanton of the size $\lambda$.
\par
Now consider the asymptotic limit $|x|\gg R$ (or, equivalently, $r\gg R$). In this case,
\begin{equation}
\rho(x)=1+\frac{\lambda^2}{2rR}+O\left(e^{-r/R}\right).
\end{equation}
Substituting this expression in (\ref{13}), we obtain
\begin{equation}
A_m\sim\frac{\lambda^2R_{mn}x^n}{4Rr^3(1+\lambda^2/2Rr)},
\end{equation}
where $n\ne 0$. Hence it follows that $A_m\sim 1/r^2$ as $r\gg \lambda^2/2R$, so the field strength falls off as $1/r^3$. If $r\ll\lambda^2/2R$, then $A_m\sim 1/r$ and hence the field strength falls off as $1/r^2$. In this case, the space-time components of the field strength looks like that of a monopole in seven dimensions.

\section{Conclusion}

In this paper, we have studied the 't Hooft type instantons in eight dimensions, which satisfy the (anti)self-dual equations $\ast F\wedge F=\pm F\wedge F$. Special attention was paid to the study of the configuration of two self-dual instantons. We used such instantons to construct soliton solutions for the low-energy effective theory of the heterotic fivebrane. These solutions are a generalization of the soliton solutions that were obtained in the case of one-instanton.
\par
In order to see the self-dual instanton effects of the obtained solutions, we have investigated the $\alpha'$ corrections to the eight-dimensional Yang-Mills theory which was naturally realized as a low-energy effective field theory. We have shown that the self-dual two-instanton configurations satisfy the $D$-instanton conditions in the zero-size and can be identified with the $D$-instantons embedded in the $D7$-brane world volume.
\par
Along with the study of the self-dual two-instantons, new solutions of the antiself-duality equations in $4n$ dimensions was found. It is curious that the topological charge of this antiself-dual instanton, determined by the fourth Chern number, turned out to be zero. For this reason, using this instanton in the low-energy effective theory did not lead to the appearance of a new fivebrane solution. We got the elementary string solution. In addition, we have constructed an infinite series of exotic antiself-dual instantons in $4n$ dimensions.
\par
In the last section, we studied eight-dimensional periodic instantons (calorons) and their relationship with monopoles. Using the standard construction of calorons, we have found solutions to the self-duality equations on the space $R^7\times S^1$. Investigating this instanton solution in various asymptotic limits, we constructed static monopole seven-dimensional solutions.

\end{document}